\documentclass[mathleft
]{an}
\usepackage{graphicx}
\usepackage{times}
\begin{document}

\Pagespan{789}{}
\Yearpublication{2006}%
\Yearsubmission{2005}%
\Month{11}%
\Volume{999}%
\Issue{88}%

\title{Inpainting: A powerful interpolation technique for helio- and asteroseismic data}

\author{K.H. Sato\inst{1}\fnmsep\thanks{Corresponding author:
  \email{kumiko.sato@cea.fr}\newline}
\and R.A. Garc\'\i a\inst{1}
\and S. Pires\inst{1}
\and J. Ballot\inst{2}
\and S. Mathur\inst{3}
\and B. Mosser\inst{4}
\and E. Rodriguez\inst{5}
\and J.L. Starck\inst{1}
\and K. Uytterhoeven\inst{1}
}
\titlerunning{Inpainting}
\authorrunning{K.H. Sato et al.}
\institute{
Laboratoire AIM, CEA/DSM-CNRS, Universit\'e Paris 7 Diderot, IRFU/SAp-SEDI, Centre de Saclay, 91191, Gif-sur-Yvette, France
\and 
Laboratoire d'Astrophysique de Toulouse-Tarbes, Universit\'e de Toulouse, CNRS, F-31400, Toulouse, France
\and 
High Altitude Observatory, Boulder, CO, 80302, USA
\and
LESIA, UMR8109, Universit\'e Pierre et Marie Curie, Universit\'e Denis Diderot, Obs. de Paris, 92195 Meudon Cedex, France
\and
Instituto de Astrof\'\i sica de Andaluc\'\i a (CSIC), Apartado 3004, 18080 Granada, Spain
}

\received{}
\accepted{}
\publonline{later}


\abstract{In Helio- and asteroseismology, it is important to have continuous, uninterrupted, data sets. However, seismic observations usually contain gaps and we need to take them into account. In particular, if the gaps are not randomly distributed, they will produce a peak and a series of harmonics in the periodogram that will destroy the stellar information. An interpolation of the data can be good solution for this problem.
In this paper we have studied an interpolation method based on the so-called 'inpainting' algorithms. To check the algorithm, we used both VIRGO and CoRoT satellite data to which we applied a realistic artificial window of a real CoRoT observing run to introduce gaps. Next we compared the results with the original, non-windowed data. Therefore, we were able to optimize the algorithm by minimizing the difference between the power spectrum density of the data with gaps and the complete time series. In general, we find that the power spectrum of the inpainted time series is very similar to the original, unperturbed one. Seismic inferences obtained after interpolating the data are the same as in the original case.}

\maketitle

\section{Introduction}

Helio- and asteroseismology are power tools to accurately determine the structure of the stellar interiors (e.g. Chris\-ten\-sen-Dalsgaard et al. 1996; Chaplin et al. 2008), their dynamics (Thompson et al. 1996; Garc\'\i a et al. 2008) as well as global parameters as their masses, radius and ages (e.g. Stello et al. 2009).

To do so,  it is important to have continuous data without regular gaps that would introduce a series of spurious peaks in the power spectrum (e.g. Mosser et al. 2008). For instance, the time series obtained with the observations of the CoRoT (Convection, Rotation and planetary Transits) satellite (Michel et al. 2008) are periodically perturbed by high-energy particles hitting the satellite when it is crossing the South Atlantic Anomaly (SAA) (e.g. Auvergne et al. 2009). The presence of repetitive gaps, which come from this regular perturbation, induces spurious peak in the power spectrum. To reduce the influence of these non-desirable peaks, it is commonly used to interpolate the data. In some cases, a linear interpolation is sufficient to do so (e.g. Appourchaux et al. 2008; Benomar et al 2009; Garc\'\i a et al. 2009, Deheuvels et al. 2010) but in other cases a more sophisticated algorithm is necessary (e.g. Mosser et al. 2009). In this paper, we propose a different algorithm based on the so-called {\it inpainting} techniques (Elad et al. 2005; Pires et al. 2009) that seems to be especially suited for our purposes. All improvements in the gap-filling data are of special importance for the analysis of CoRoT data but also for the forthcoming Kepler observations, for which very long time series (more than 3.5 years) are being expected for thousands different stars covering the HR diagram (e.g. Bedding et al. 2010; Chaplin et al. 2010; Stello et al. 2010).

\section{Inpainting algorithm}
Inpainting techniques are known in the field of image processing. The method that is used in this paper relies on the sparse representation of the data introduced by Elad et al. (2005). It assumes that there exists a dictionary $\Phi$ where the complete data are sparse and the incomplete data are less sparse. It means that there exists a representation $\alpha = \Phi^TX$ of the signal $X$ in the dictionary $\Phi$ where most coefficients $\alpha_i$ are close to zero.

The solution is obtained by minimizing the following equation:
\begin{equation}
\label{eq.1}
\min_{X}  \| \Phi^T X \|_1    \textrm{ subject to }   \sum_i (Y - MX)^2 \le \sigma,
\label{eq1}
\end{equation}  
whereby, $X$ is the ideal complete time series, $Y$ the observed time series and $M$ the mask (i.e. $M_i = 1$ is a valid data point and $M_i=0$ elsewhere). Inpainting consists in recovering $X$ knowing $Y$ and $M$. 
In equation (\ref{eq1}), $\sigma$ stands for the noise standard deviation and we use a pseudo norm with $ \| z \|_1 = \sum_i z_i$. 

\begin{figure}[htbp]
\includegraphics[width=7.5cm, trim= 0mm 0mm 0mm 67mm, clip]{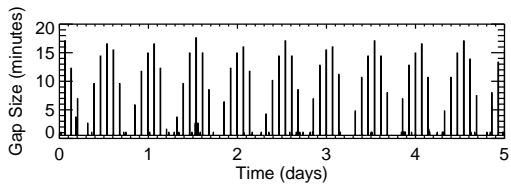}
\caption{Sample of the first 5 days of a typical CoRoT observation mask  corresponding to the LRc02 observations.}
\label{gap}
\end{figure}

In helio- and asteroseismology the best dictionary is ba\-sed on Discrete Cosinus Transforms (DCT). In the case of CoRoT, we have built the mask $M$ to remove those data points that were affected by the SAA crossing and also all the other points that were flagged in the datasets as bad points, according to the status flag (Auvergne et al. 2009). In Fig.\ref{gap}, we show a sample of a typical CoRoT observation mask. The masked gaps in the CoRoT time series typically have time scales less than 20 minutes and periodic patterns that originate from the orbital period of the satellite. About 10\% of the data points are flagged as bad. Sometimes, longer time gaps of the order of one hour occur in the CoRoT data. To treat the large variation of gap sizes, we used a wavelet decomposition and we determined the range of frequencies that we can interpolate in each gap by changing the blocksize of the local DCT for each wavelet plane. This corresponds to a  {\it Multi Scale Discrete Cosinus Transform} (MSDCT).

\section{Test of the inpainting algorithm}
\subsection{Test with VIRGO data}
We tested our inpainting algorithm on the VIRGO/SOHO data (Fr{\"o}hlich et al. 1995) by applying the observational mask of CoRoT observations with a typical duty cycle of $90\%$. With this test, we optimized the algorithm to minimize the difference between the power spectrum density (PSD) of the gapped time series compared to the original ones. We also applied a linear interpolation on the masked time series to check for improvement. 

\begin{figure}[!htbp]
\includegraphics[angle=90,width=8cm, trim=0mm 0mm 20mm 10mm]{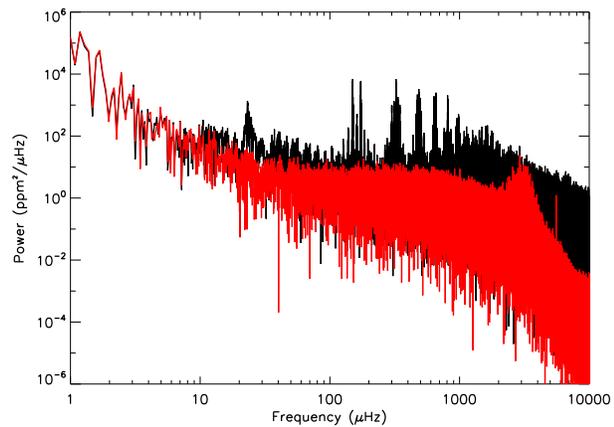}
\caption{PSD (in units of ${\rm ppm^2/\mu Hz}$) of the original (red) and masked (black) time series of VIRGO data.}
\label{VIRGO_PSD_1}
\end{figure}

\begin{figure}[!htbp]
\includegraphics[angle=90,width=8cm, trim=0mm 0mm 35mm 20mm]{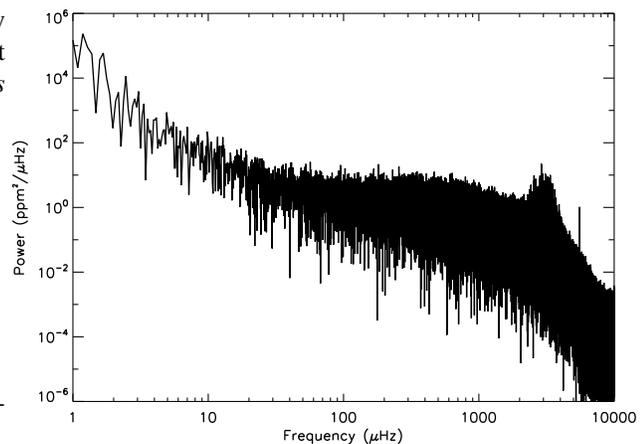}
\caption{PSD (in units of ${\rm ppm^2/\mu Hz}$) of the inpainted time series of VIRGO data. }
\label{VIRGO_PSD_2}
\end{figure}

Fig.\ref{VIRGO_PSD_1} shows the PSD of the original VIRGO data (red) and the masked time series (black). 
The PSD of the gapped time series has a peak at 161.7$\mu$Hz (orbital frequency) and a sequence of harmonics combined with daily aliases ($\pm$ 11.57 $\cdot k$ $\mu$Hz, where $k$ is an interger). This pattern of spurious peaks makes it difficult to find and identify the p-mode signature at 3 mHz.
We then applied our inpainting algorithm to this gapped time series. The resulting PSD is shown in Fig.\ref{VIRGO_PSD_2}. It is clear that the inpainting algorithm reduced the non-desirable sequence of peaks due to the gaps. 


\begin{figure}[!htbp]
\includegraphics[angle=90,height=4cm,width=7.5cm]{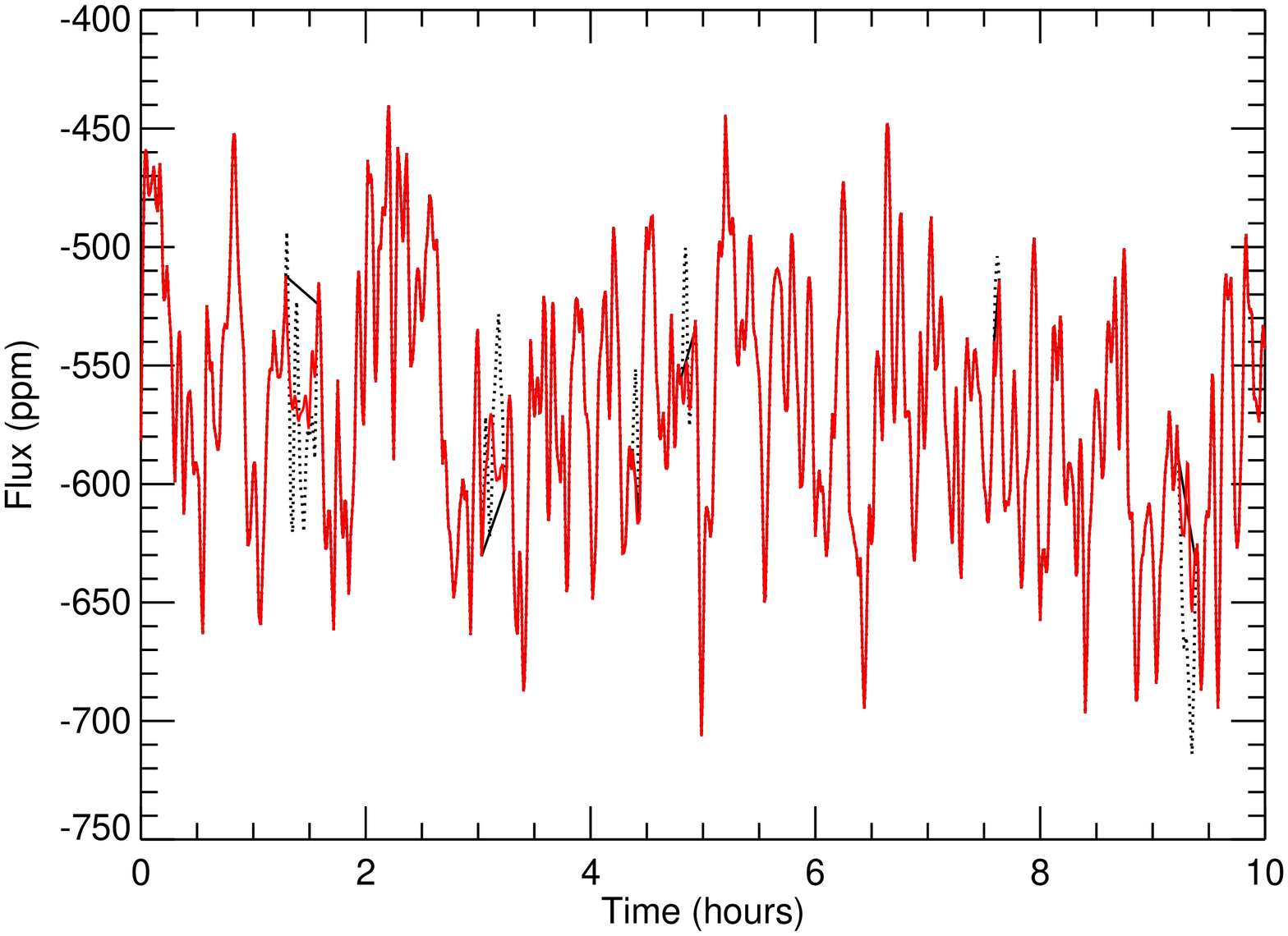}
\includegraphics[angle=90,height=4cm,width=7.5cm]{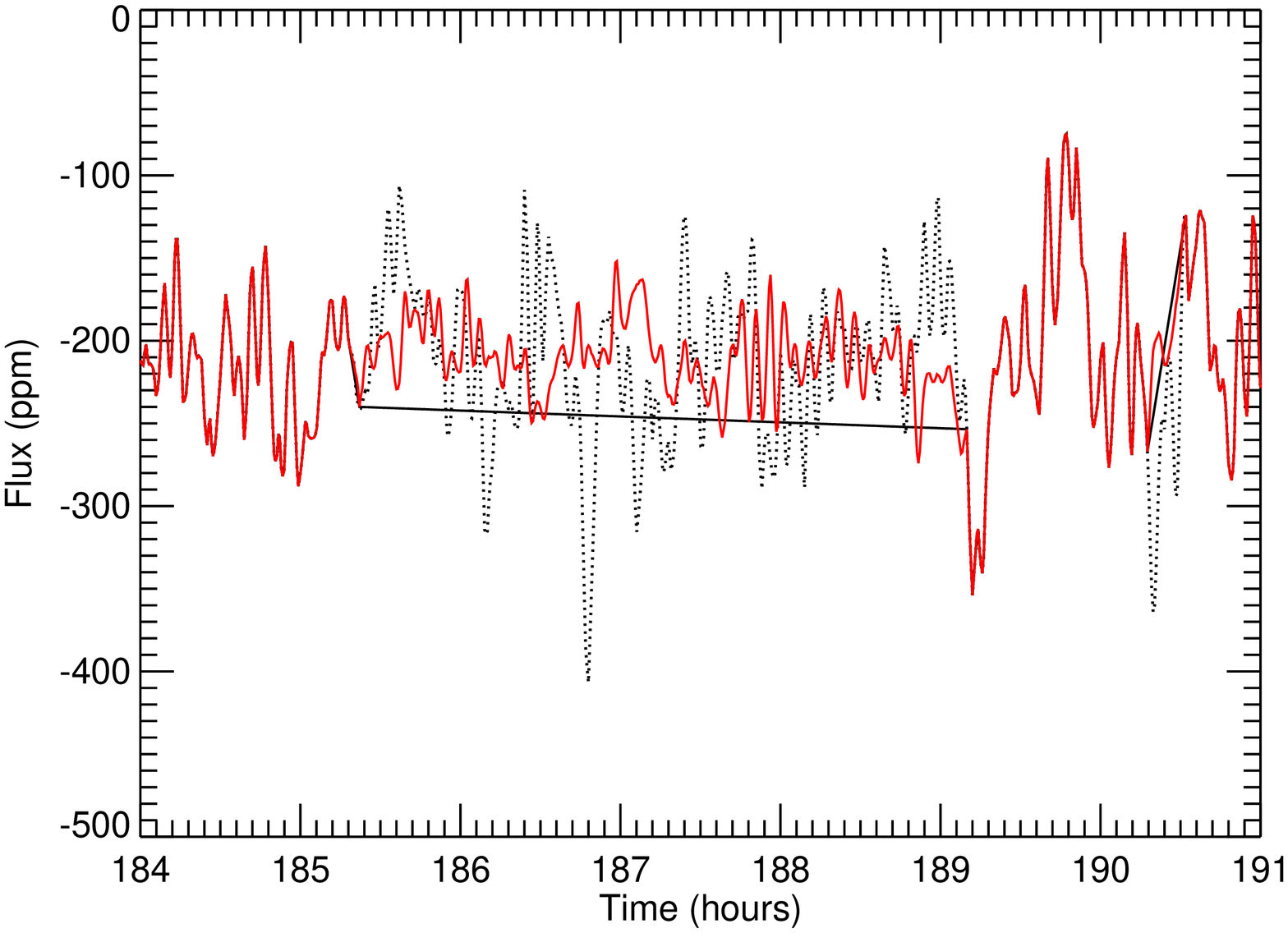}
\caption{Sample of the time series with the original VIRGO data (dotted line), inpainted data (red solid line) and linear interpolated data (black solid line).}
\label{VIRGO_curve}
\end{figure}

Examples of the inpainted (red solid line) and the linearly interpolated (black solid line) time series are illustrated in Fig.\ref{VIRGO_curve}. 
The inpainting method describes data in the gaps of the time series thanks to an extrapolation based on the frequency content as derived from DCT. This is a main feature of the inpainting algorithm. In other words, the inpainting algorithm tries to reconstruct the data inside the gaps from the available data.

\begin{figure}
\includegraphics[angle=90,width=7cm]{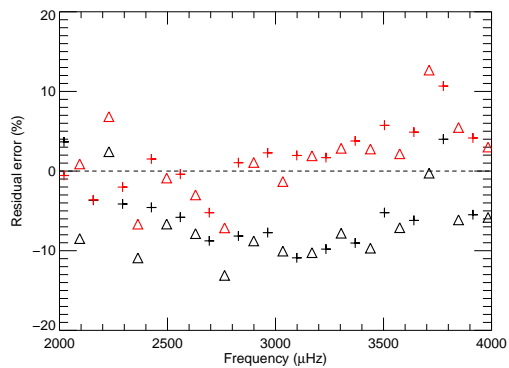}
\caption{Fractional difference in power between the original and the inpainted time series of VIRGO data for some $l=0$ (triangle) and $l=1$ (plus) modes. Red symbols and black symbols correspond to inpainting and linear interpolation respectively.}
\label{VIRGO_error}
\end{figure}

To check the amplitudes of the modes in the inpainted PSD, we have calculated  the fractional difference in power between the original and the interpolated series for $l=0$ and $l=1$ mode frequencies (see Fig.\ref{VIRGO_error}). The amplitudes of the modes in the linearly interpolated series were underestimated by 10$\%$ at the maximum of the p-mode hump and they depend on frequency. In the other hand, the inpainting retrieved amplitudes are roughly the same than in the original series but with some excess in power above 3.6 mHz that is currently under study. 

\subsection{Test on CoRoT data}
We have also tested our inpainting algorithm on the initial CoRoT run in which the "HELREG" calibrated data (see for the details Auvergne et al. 2009) produces a PSD with only a few orbital harmonics. For this test, we chose two different type of pulsating stars: a solar-like star with solar-like oscillations and a $\gamma$ Doradus star. 

\paragraph{Solar-like star:}
We tested the inpainting technique on the CoRoT target HD 181420 (Barban et al. 2009). In this run, the standard linear interpolation algorithm removed ne\-arly all the harmonics of the orbital period of 161.7 $\mu$Hz and an optimized interpolation technique was not needed. Thus, we multiplied this light curve by the observational window  of the second run in the galactic center direction (LRc02) 
for which the linear interpolation could not clean up the resultant PSD (see Sect. 4), and we ran the same tests as in the case of the VIRGO data. The PSD of the inpainted  light curve is less noisy than the masked, linearly interpolated one and the characteristics of the p-modes retrieved from the inpainted data are similar to the ones retrieved from the original data set within the error bars. 

\begin{figure}
\includegraphics[angle=-90,width=7cm]{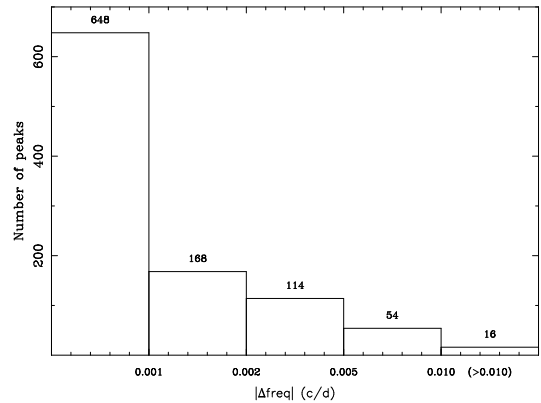}
\caption{Distribution of frequency differences between frequencies detected from the original (HELREG) and the inpainted time series.}
\label{hd49434}
\end{figure}

\paragraph{$\gamma$ Doradus star:}
We also tested the inpainting method on the CoRoT data of HD 49434 (Uytterhoeven et al. 2008; Chapellier et al. 2010) observed on the anti-center galactic direction. In this run, it was also not necessary to use an optimized interpolation technique to obtain a PSD almost free of orbital perturbations. Thus, we have again multiplied this light curve by the observational window of the LRc02 run. Like in the solar-like star, the inpainted PSD is less noisy than the original with the LRc02 window.
Then, to check if there was an influence on the asteroseismic results, we performed a frequency analysis on the original and the inpainted PSD. Fig.\ref{hd49434} shows the distribution of frequency differences between frequencies detected in the original and the inpainted time series. The frequency resolution of the time series is  $\approx 0.0066 \rm{c/d}$ ($\sim 0.08 \mu$Hz). That means that most of the frequencies detected in the inpainted data coincide within $\pm$ one resolution bin.   

\section{Applying the inpainting interpolation to stars in the LRc02 run}
The stars observed in the LRc02 CoRoT field suffer from a pollution in the PSD with a sequence of orbital harmonics plus some daily aliases around each orbital peak. We applied the inpainting algorithm on the problematic LRc02 CoRoT time series of two main target : HD\ 170987 (Mathur et al. 2010a) and HD\ 171834 (Uytterhoeven et al. 2010).  

\begin{figure}
\includegraphics[angle=90,height= 4cm, width=7.5cm, trim=0mm 0mm 16mm 10mm]{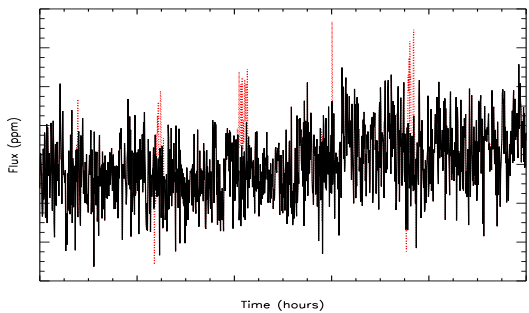}
\caption{Sample of the raw (HELREG) (red) and inpainted (black) time series of HD\ 170987.}
\label{HD170987_curve}
\end{figure}

\begin{figure}
\includegraphics[angle=90,width=7.5cm, trim=5mm 0mm 16mm 10mm]{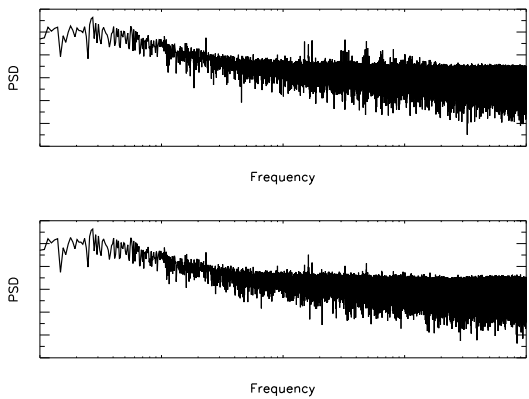}
\caption{PSD of the raw (HELREG) (top) and inpainting (bottom) time series of HD\ 170987.}
\label{HD170987_PSD}
\end{figure}

Fig.\ref{HD170987_curve} shows a sample of the raw and inpainted time series of HD\ 170987, a solar-like star.
The top panel of Fig.\ref{HD170987_PSD} shows the PSD of the raw time series of this star. The sequence of the orbital frequency and its daily aliases are clearly polluting the full spectrum.
The bottom panel of Fig.\ref{HD170987_PSD} shows the PSD of the inpainted time series. Only the first harmonics of the orbit are visible.

\begin{figure}
\includegraphics[angle=90,width=7.5cm, trim=5mm 0mm 16mm 10mm]{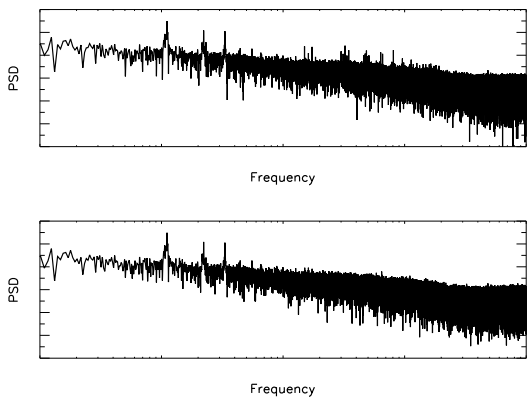}
\caption{PSD of the raw (HELREG) (top) and inpainted (bottom) time series of HD\ 171834.}
\label{HD171834_PSD}
\end{figure}

Fig.\ref{HD171834_PSD} shows the PSD of the raw (HELREG) and inpainted time series of HD\ 171834. Once again, the orbital harmonics and the daily aliases are significantly reduced after inpainting improving the seismic analysis of this star.

\section{Conclusions}
We have shown that the inpainting based on MSDCT is a powerful interpolation algorithm which is well adapted to correct the data gaps in helio and asteroseismic observations. We already applied it to the CoRoT data of the solar-like target HD\ 170987 (Mathur et al. 2010a). We are planning to integrate it into our asteroseismic automatic pipeline for the analysis of Kepler data (Mathur et al. 2010b), and to use it to correct the GOLF velocity time series (Garc\'\i a et al. 2005). 

\acknowledgements
CoRoT was developed by the French Space agency CNES in collaboration with the Science Programs of ESA, Austria, Belgium, Brazil, Germany and Spain. GOLF and VIRGO instrument onboard SOHO are a cooperative effort of many individuals, to whom we are indebted. SOHO is a project of international collaboration between ESA and NASA. This work has been partially funded by the GOLF/CNES grant at the CEA/Saclay.


\end{document}